\documentclass{PoS}
\usepackage{amsmath,amssymb}
\usepackage{grffile}
\title{Can the complex Langevin method see 
the deconfinement phase transition in QCD at finite density?}

\ShortTitle{Can the CLM see the deconfinement phase transition 
in QCD at finite density?}

\author{\speaker{Shoichiro Tsutsui}\thanks{KEK-TH/2086}\\
        High Energy Accelerator Research Organization (KEK),\\
        1-1 Oho, Tsukuba, Ibaraki 305-0801, Japan\\
        E-mail: \email{stsutsui@post.kek.jp}}

\author{Yuta Ito\\
        High Energy Accelerator Research Organization (KEK),\\
        1-1 Oho, Tsukuba, Ibaraki 305-0801, Japan\\
	E-mail: \email{yito@post.kek.jp}}

\author{Hideo Matsufuru\\
        High Energy Accelerator Research Organization (KEK),\\
	1-1 Oho, Tsukuba, Ibaraki 305-0801, Japan\\
	E-mail: \email{hideo.matsufuru@kek.jp}}

\author{Jun Nishimura\\
        High Energy Accelerator Research Organization (KEK),\\
	1-1 Oho, Tsukuba, Ibaraki 305-0801, Japan, and\\
	Graduate University for Advanced Studies (SOKENDAI),\\
        1-1 Oho, Tsukuba 305-0801, Japan\\
	E-mail: \email{jnishi@post.kek.jp}}

\author{Shinji Shimasaki\\
	Research and Education Center for Natural Sciences,
	Keio University,\\
        Hiyoshi 4-1-1, Yokohama, Kanagawa 223-8521, Japan\\
	E-mail: \email{shinji.shimasaki@keio.jp}}

\author{Asato Tsuchiya\\
	Department of Physics, Shizuoka University,
	836 Ohya, Suruga-ku, Shizuoka 422-8529, Japan\\
	E-mail: \email{tsuchiya.asato@shizuoka.ac.jp}}

\abstract{Exploring the phase diagram of QCD 
at finite density is a challenging problem since 
first-principle calculations based on standard Monte Carlo methods 
suffer from the sign problem. As a promising approach to this issue, 
the complex Langevin method (CLM) has been pursued intensively.
In this work, we investigate the applicability of the CLM in the vicinity 
of the deconfinement phase transition using the four-flavor staggered fermions. 
In particular, we look for a signal of the expected first order phase transition
within the validity region of the CLM.
}

\FullConference{The 36th Annual International Symposium on Lattice Field Theory
- LATTICE2018\\
22-28 July, 2018\\
Michigan State University, East Lansing, Michigan, USA.}

\begin{document}

\section{Introduction}
Exploring the phase diagram of QCD at finite density and temperature
is important due to its relevance
to the heavy-ion collision
physics and the determination of the equation of state for neutron stars.
However,
lattice QCD simulations based on conventional Monte Carlo algorithms 
suffer from a severe sign problem in the finite density region.

To overcome this problem,
the complex Langevin method (CLM)~\cite{Parisi:1984cs,Klauder:1983sp} has 
been investigated intensively in recent years.
Basically, 
it
is an extension of the stochastic quantization 
to theories with a complex action.
In this framework, 
the expectation value of holomorphic observables is computed by 
solving the complex Langevin equation, which describes the stochastic 
time-evolution 
of the complexified dynamical variables.
This procedure does not rely on the probabilistic interpretation 
of the Boltzmann weight $e^{-S}$,
and hence it is free from the sign problem.
However, the equivalence to the familiar path integral quantization
does not always hold \cite{Aarts:2009uq,Aarts:2011ax}
unlike the case with a real action.
Recently, a necessary and sufficient condition for the equivalence
based on the probability distribution of the drift term
has been proposed~\cite{Nagata:2016vkn}; i.e.,
the CLM gives correct results
if and only if the probability distribution of the drift term decays 
exponentially or faster.
Because of this condition, the CLM works in some parameter region of
finite density QCD, but not in the entire region.
As we will see in the following, 
the distribution falls off with a power law in some region,
which implies that the CLM is no longer valid there.
Since one can easily monitor the distribution while generating configurations,
it is useful and preferable to judge the validity of the CLM 
in this way in actual simulations for each set of parameters.

In this paper we discuss the applicability of the CLM in the vicinity 
of the deconfinement transition with $N_{\rm f}=4$ 
staggered fermions at finite temperature $T$ and finite chemical potential $\mu$.
This transition is known to be of 
first order at $\mu = 0$ \cite{Fukugita:1990vu},
and it is expected to be so also at $\mu\neq0$
based on the canonical ensemble method \cite{deForcrand:2006ec,Li:2010qf}.
Recently the CLM and the standard reweighting method have been applied 
to this theory for comparison \cite{Fodor:2015doa}.
While the deconfinement transition 
was accessible by the reweighting method unless $\mu$ or the lattice size
is not too large,
it turned out to be difficult to access by the CLM
for the chosen setup because
the simulation becomes unstable for 
small $\beta$.
Motivated by this result, we perform simulations with larger lattice size
in the temporal direction so that the phase transition occurs at larger $\beta$,
and investigate whether the CLM has an ability to probe
the deconfinement transition in different setups.

The rest of this article is organized as follows.
In section~\ref{method} we give a brief overview 
on the application of the CLM to lattice QCD at finite density. 
In section~\ref{result} we present our results.
Section~\ref{summary} is devoted to a summary and discussions.

\section{Complex Langevin method for finite density QCD}\label{method}
We apply the CLM to lattice QCD on a four-dimensional lattice 
with the temporal extent $N_{\rm t}$ and the spatial extent $N_{\rm s}$.
Throughout this paper, we set the lattice spacing to unity.
The partition function is given by
\begin{align}
Z = \int \prod_{x\nu}dU_{x\nu} \,
\mathrm{det}M(U;\mu) \, e^{-S_{\rm g}(U)} \ ,
\label{Z}
\end{align}
where $U_{x\nu}\in \mathrm{SU}(3)$ ($\nu=1,2,3,4$) are the link variables
with $x=(x_1,x_2,x_3,x_4)$ being the coordinates of each site.
For the gauge action $S_{\rm g}(U)$, 
we use the Wilson plaquette action defined by
\begin{align}
S_{\rm g}
= -\frac{\beta}{6}
\sum_{x}\sum_{\mu <\nu}\mathrm{tr}
\Big(U_{x,\mu\nu}+U_{x,\mu\nu}^{-1}\Big) \ , 
\quad\quad
U_{x,\mu\nu} = U_{x\mu}U_{x+\hat\mu,\nu}U_{x+\hat\nu,\mu}^{-1}U_{x\nu}^{-1} \ ,
\end{align}
where $\hat\mu$ is the unit vector in the $\mu$ direction.
For fermions, 
we use the $N_{\rm f}=4$ unimproved staggered fermion 
with the degenerate quark mass $m$,
which corresponds to choosing the fermion matrix $M(U;\mu)$ in (\ref{Z}) as
\begin{align}
M(U;\mu)_{xy}
=
m\delta_{xy}
+ \sum_{\nu=1}^{4} \frac{\eta_\nu(x)}{2}
\left(e^{\mu \delta_{\nu 4}}U_{x\nu}\delta_{x+\hat\nu , y}
-e^{-\mu \delta_{\nu 4}}U_{x-\hat\nu , \nu}^{-1}
\delta_{x-\hat\nu , y}\right) \ ,
\label{fermion matrix}
\end{align}
where $\mu$ represents the quark chemical potential
and $\eta_\nu(x) = (-1)^{x_1+\cdots+x_{\nu-1}}$ is 
a site-dependent sign factor.
The sign problem is caused by the fermion determinant $\mathrm{det}M(U;\mu)$
appearing in (\ref{Z}).
We impose anti-periodic boundary conditions for the fermionic field
in the temporal direction. The temperature is then given by $T=1/N_{\rm t}$.

The CLM is one of the most promising approaches to overcome the sign problem.
In this method, 
the link variables $U_{x\mu}$
defined as $\mathrm{SU}(3)$ matrices are complexified into 
$\mathrm{SL}(3,\mathbb{C})$ matrices, which we denote 
as $\mathcal U_{x\mu}$.
Then we consider a fictitious time evolution of
the complexified variables given by the complex Langevin equation
with stepsize $\epsilon$
\begin{align}
\mathcal U_{x\mu}(t+\epsilon)
=\exp\left(i\sum_{a=1}^{8}\lambda_a
\left[-\epsilon v_{ax\mu}(\mathcal U(t))
+\sqrt{\epsilon}\eta_{ax\mu}(t)\right]  \right) \ 
\mathcal U_{x\mu}(t) \ ,
\label{CLM}
\end{align}
where $\lambda_a \ (a=1,\cdots,8)$ are
the SU(3) generators normalized by $\mathrm{tr}(\lambda_a \lambda_b) = \delta_{ab}$.
The noise term $\eta_{ax\mu}(t)$ is composed of real gaussian random numbers
normalized as 
\begin{align}
\langle \eta_{ax\mu}(s) \eta_{b y\nu}(t) \rangle = 
2 \delta_{ab} \delta_{xy} \delta_{\mu\nu}  \delta_{st} \ .
\end{align}
The drift term $v_{ax\mu}(\mathcal U(t))$
is defined
by the holomorphic extension of
\begin{align}
 v_{ax\mu}(U)
= \left.\frac{d}{d\alpha}S(e^{i \alpha \lambda_a}U_{x\mu})\right|_{\alpha=0}
\label{drift}
\end{align}
defined for the SU(3) link variables 
with the total action $S[U] = S_{\rm g}[U] - \log \det M(U; \mu)$.

A subtle point of the CLM is that it is not guaranteed to yield correct results,
and hence one has to check the reliability of the results 
after generating configurations.
According to the criterion proposed in ref.~\cite{Nagata:2016vkn},
the CLM is equivalent to the usual path integral formulation
if the probability distribution of the drift term shows an exponential fall-off.
In the case of finite density QCD, we define the magnitude of the drift term by
\begin{align}
  v= \max _{x,\mu} \sqrt{\frac{1}{3}\sum_{a=1}^{8} |v_{ax\mu}(\mathcal U)|^2}  \ .
\label{drift-magnitude}
\end{align}

There are actually two cases in which the above criterion is violated.
One is the singular drift problem \cite{Mollgaard:2013qra,Nishimura:2015pba}, 
which occurs
because of the appearance of the inverse $M^{-1}$ in the drift term
when the fermion matrix $M(\mathcal{U}; \mu)$ has near-zero eigenvalues.
In order to detect this problem, we probe 
the contributions to the drift term 
from the fermion determinant $\det M(\mathcal{U}; \mu)$.

The other case which leads to the violation of the criterion is
the excursion problem \cite{Aarts:2009uq,Aarts:2011ax}, 
which occurs when the complexified link variables become
too far from unitary matrices.
In order to detect this problem,
it is useful to probe the unitarity norm defined by
\begin{align}
{\cal N} 
= \frac{1}{12 N_{\rm V}} \sum_{x\mu} \mathrm{tr} 
\left( \mathcal{U}_{x\mu}\mathcal{U}_{x\mu}^\dagger - {\bf 1} \right) \ ,
\end{align}
which measures the deviation of the complexified link variables
from SU(3). Here we defined $N_{\rm V}=N_{\rm t} N_{\rm s}^3$,
which represents the volume of the four-dimensional Euclidean space.


As important observables,
we consider
the Polyakov loop, which is defined by
\begin{align}
P &= \frac{1}{3 N_{\rm s}^3} 
 \sum_{\vec{x}} \mathrm{tr} 
\Big( U_{(\vec{x},1) ,4} U_{(\vec{x},2) ,4} \dots U_{(\vec{x},N_{t}) ,4} \Big)
 \ , 
\end{align}
where $\vec{x} = (x_1,x_2,x_3)$, and 
the baryon number density defined by
\begin{align}
n &=\frac{1}{3N_{\rm V}}\frac{\partial}{\partial \mu}\log Z
=\frac{1}{3N_{\rm V}}\left\langle 
\sum_{x} \frac{\eta_4(x)}{2}\mathrm{tr}
\Big( e^{\mu }M^{-1}_{x+\hat4, x}U_{x4}
+e^{-\mu }M^{-1}_{x-\hat4, x}U_{x-\hat4,4}^{-1} \Big) \right\rangle \ .
\label{baryon}
\end{align}

\section{Results}\label{result}

In the previous work~\cite{Fodor:2015doa},
the validity range of the CLM was discussed
with the maximum lattice size being $16^3\times 8$.
There it was found 
for $\mu/T=0.96$ with the quark mass $m = 0.01$, for instance,
that the CLM breaks down at $\beta \sim 5.15$, which prevented
the authors from reaching 
the transition point, which is slightly below $5.15$.
Based on this result, we employ a lattice with larger temporal size $N_{\rm t}=12$,
which is expected to shift the transition point to larger $\beta$,
hoping that the CLM can see the phase transition within 
the region of validity.
The quark chemical potential is chosen as $\mu=0.1$, which 
corresponds to $\mu/T = 1.2$.
We use an adaptive stepsize \cite{Aarts:2009dg}
with the initial value $\epsilon = 5 \times 10^{-5}$ and 
perform the gauge cooling~\cite{Seiler:2012wz}
(See refs.~\cite{Nagata:2015uga,Nagata:2016vkn} for its justification)
to minimize the unitarity norm ${\cal N}$ at each Langevin step.

\subsection{$\beta$-dependence at $m=0.01$}

\begin{figure}
	\centering
	\includegraphics[width=7.5cm]{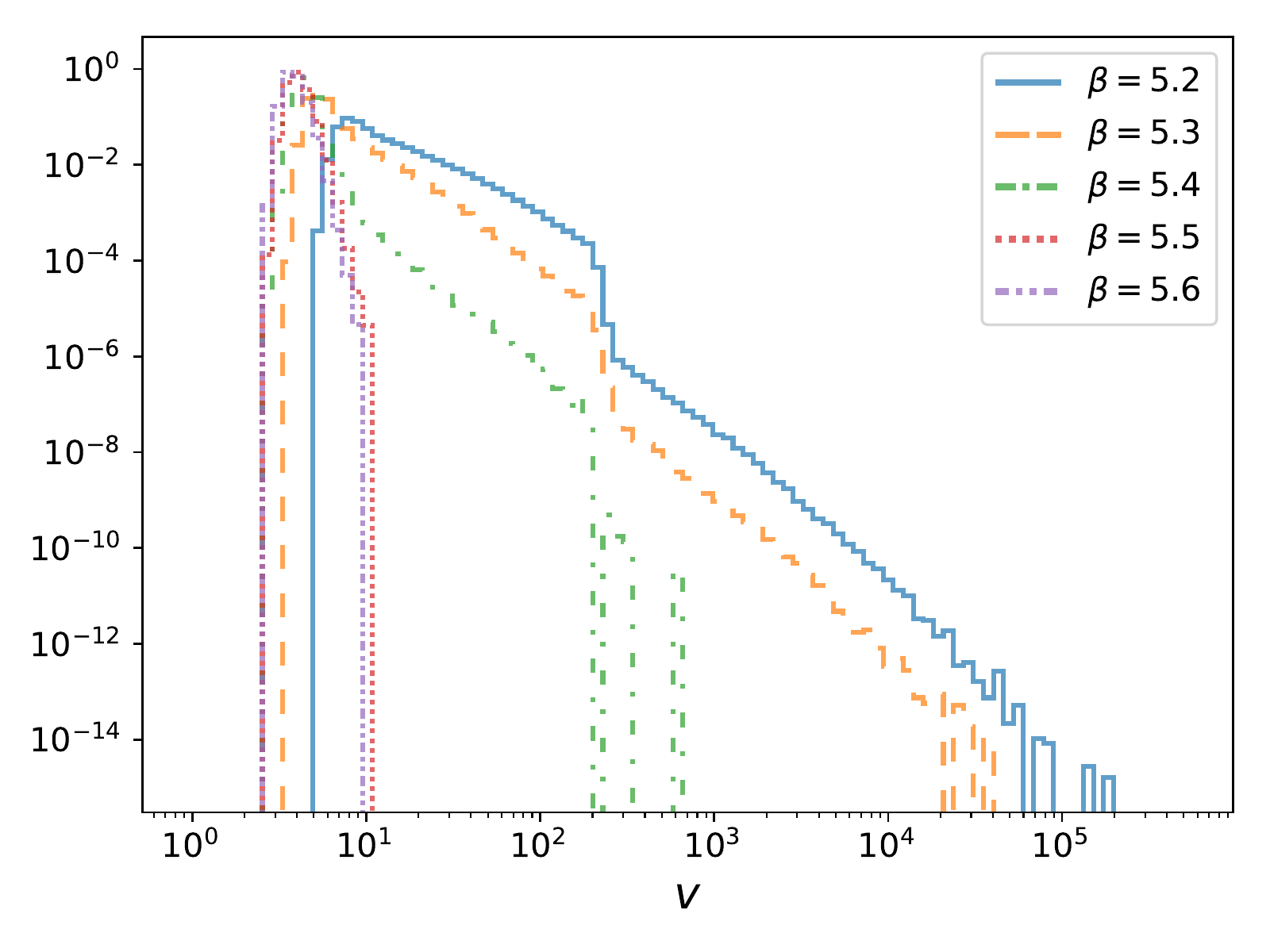}
	\includegraphics[width=7.5cm]{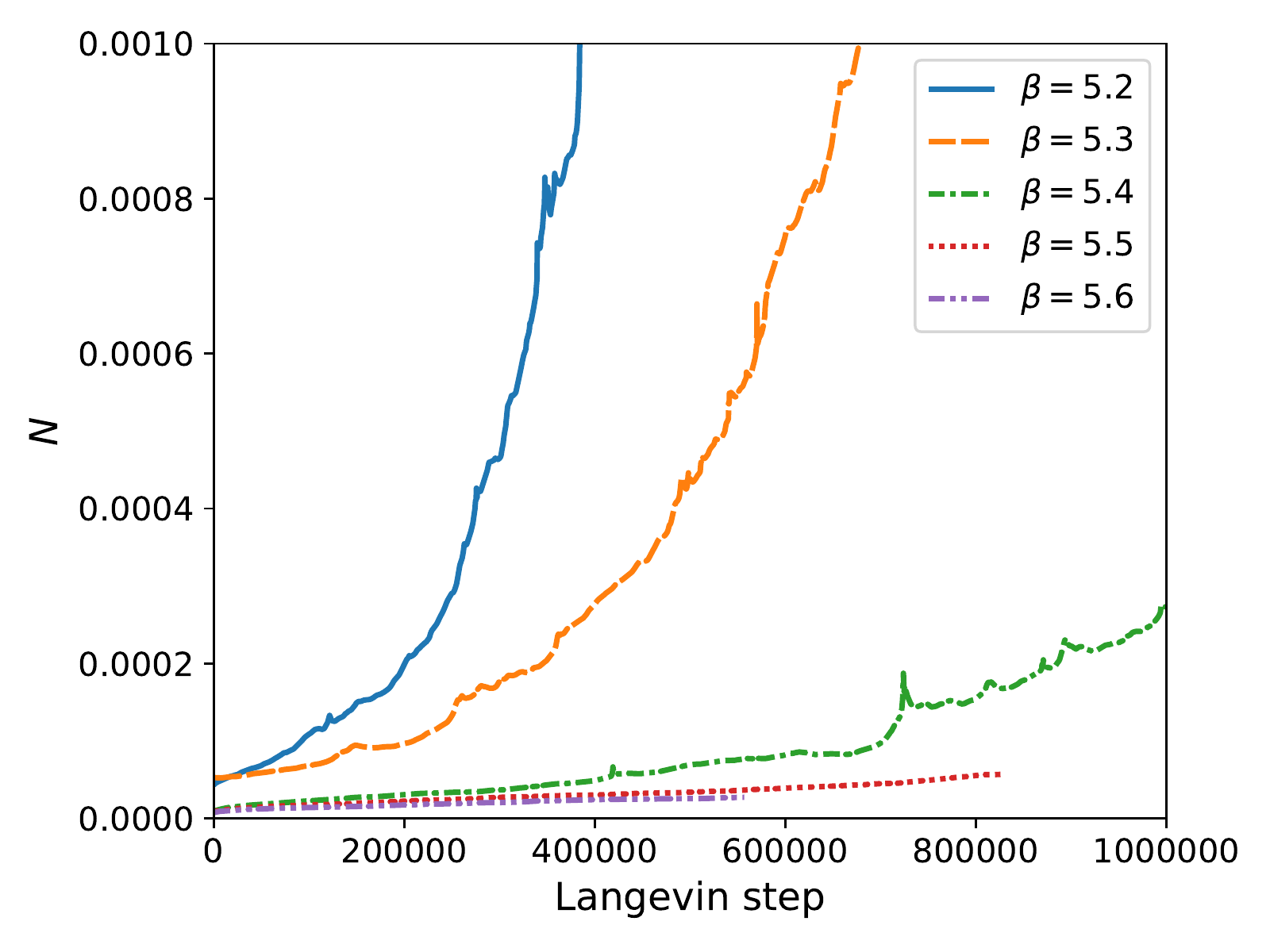}
	\caption{The histogram of the fermionic drift term (Left)
and the history of the unitarity norm (Right) are
plotted for various $5.2 \le \beta \le 5.6$ with $m=0.01$ 
on a $20^3 \times 12$ lattice.}
	\label{drift small mass}
\end{figure}

Here we show our results obtained at $m=0.01$ for $5.2 \le \beta \le 5.6$
on a $20^3 \times 12$ lattice.
First we check the reliability of the CLM.
In Fig.~\ref{drift small mass} we plot the probability distribution 
of the fermionic drift term
and the history of the unitarity norm.
%
At $\beta=5.2$ and 5.3,
we observe that the distribution of the drift term falls off with a power law
and the unitarity norm grows rapidly with the Langevin time.
Therefore, we conclude that the CLM is not reliable for these values of $\beta$.

\begin{figure}
	\centering
	\includegraphics[width=7.5cm]{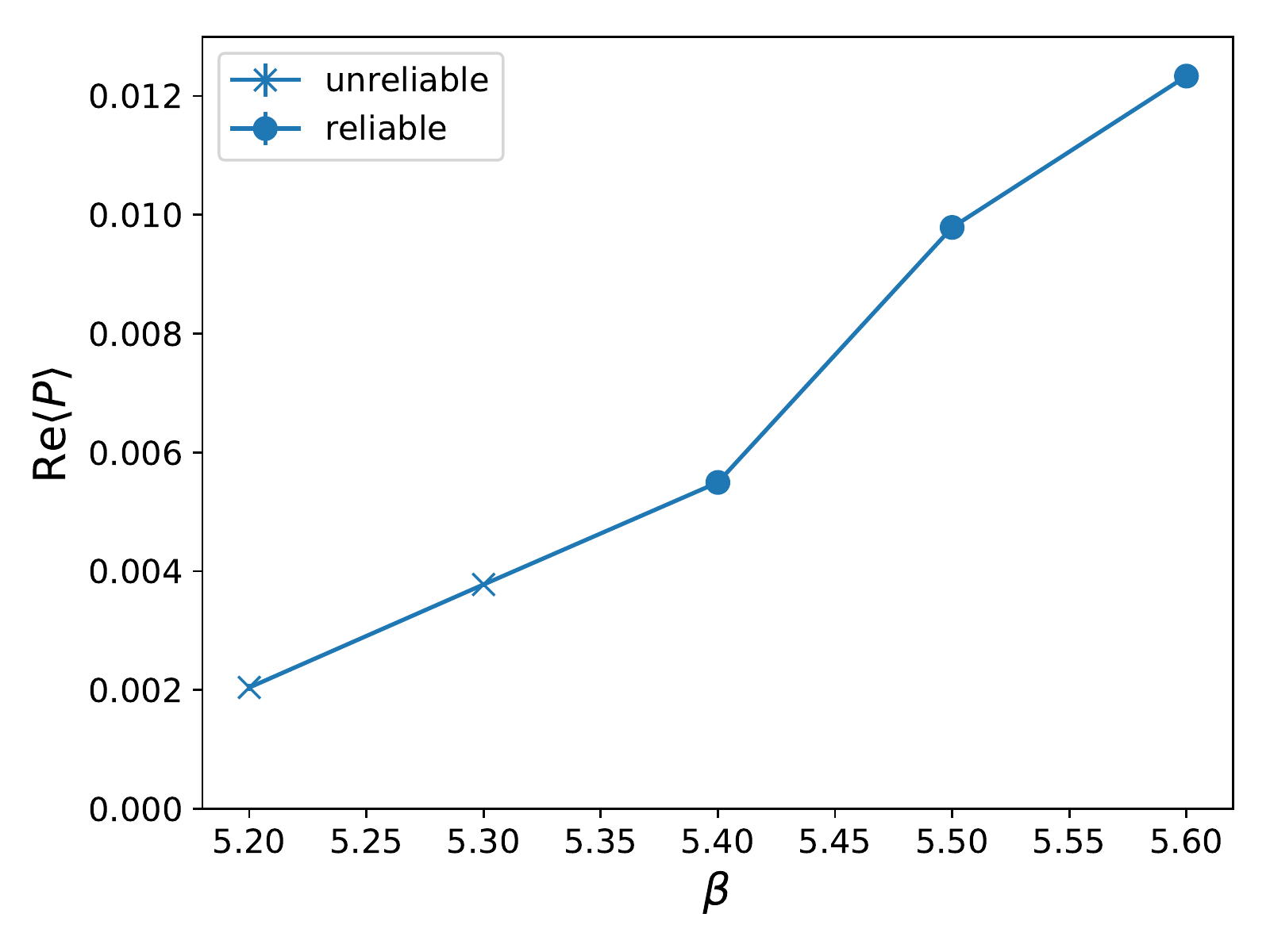}
	\includegraphics[width=7.5cm]{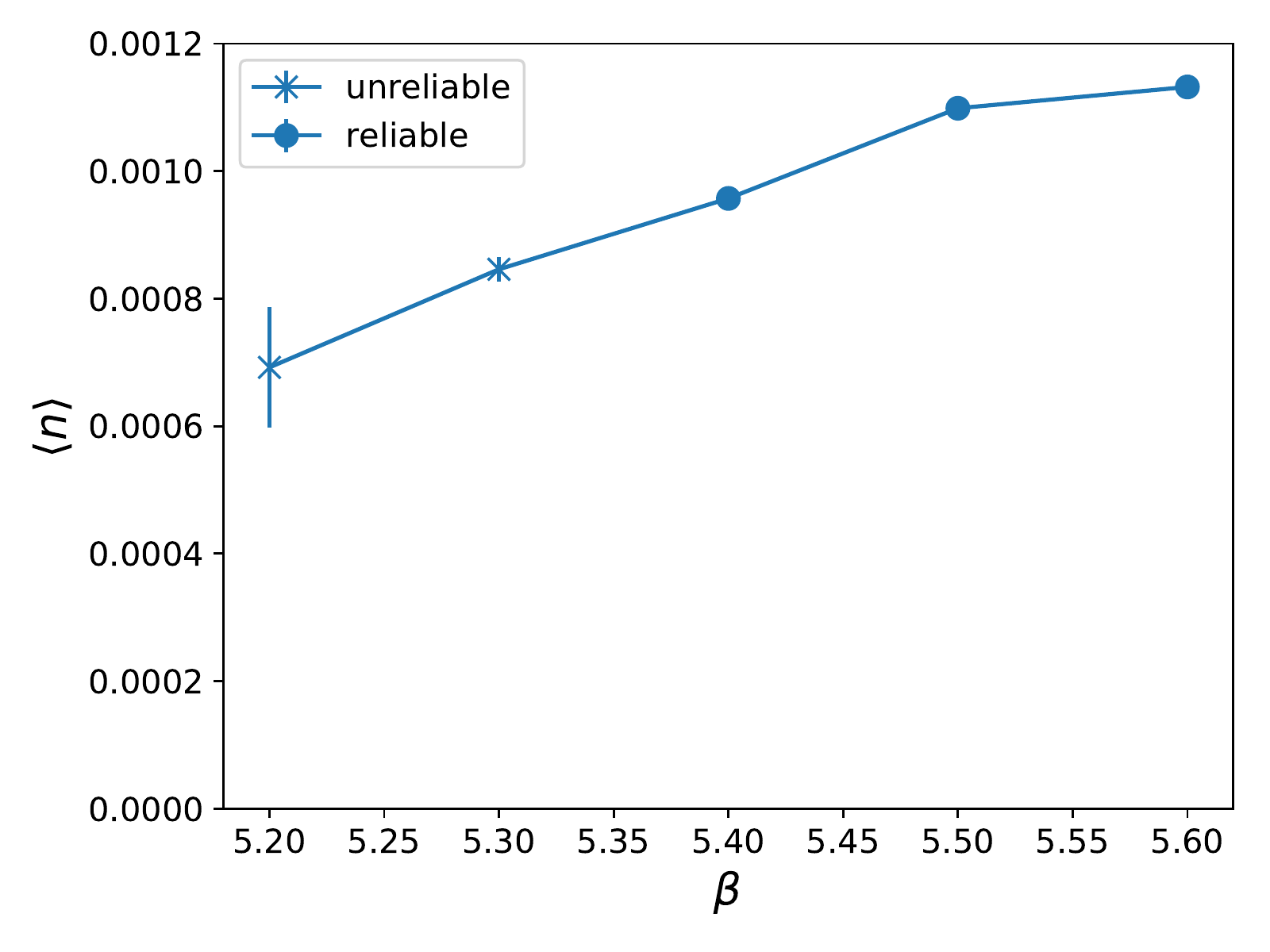}
	\caption{The real part of the Polyakov loop (Left)
and the baryon number density (Right) are plotted against $\beta$
for $m=0.01$ on a $20^3 \times 12$ lattice. 
Circles and crosses represent reliable data and unreliable data, respectively.
}
\label{obs small mass}
\end{figure}

In order to probe the deconfinement transition,
we measure the Polyakov loop and the baryon number density.
In Fig.~\ref{obs small mass}
we plot the expectation values of these quantities as a function of $\beta$.
Focusing on the reliable data at $\beta \geq 5.4$,
we find that both of them
decrease gradually as $\beta$ is lowered, but they remain
significantly away from zero suggesting
that the system is in the deconfined phase.
The first order phase transition seems to be hidden in the unreliable region,
which is similar to the situation in ref.~\cite{Fodor:2015doa}.
Thus, we find that increasing the temporal size of the lattice
does not enable us to see the deconfinement phase transition.

\subsection{Increasing the quark mass}
\label{quark_mass}

From Figs.~10 and 11 of ref.~\cite{Fodor:2015doa}, we find that
the critical $\beta$
can be shifted to larger values also by increasing the quark mass,
which provides us with another possibility to observe the phase transition
by the CLM.
Below we show our results obtained at $0.01 \le m \le 0.5$ with 
fixed $\beta=5.4$ on a $24^3 \times 12$ lattice. 

\begin{figure}
	\centering
	\includegraphics[width=7.5cm]{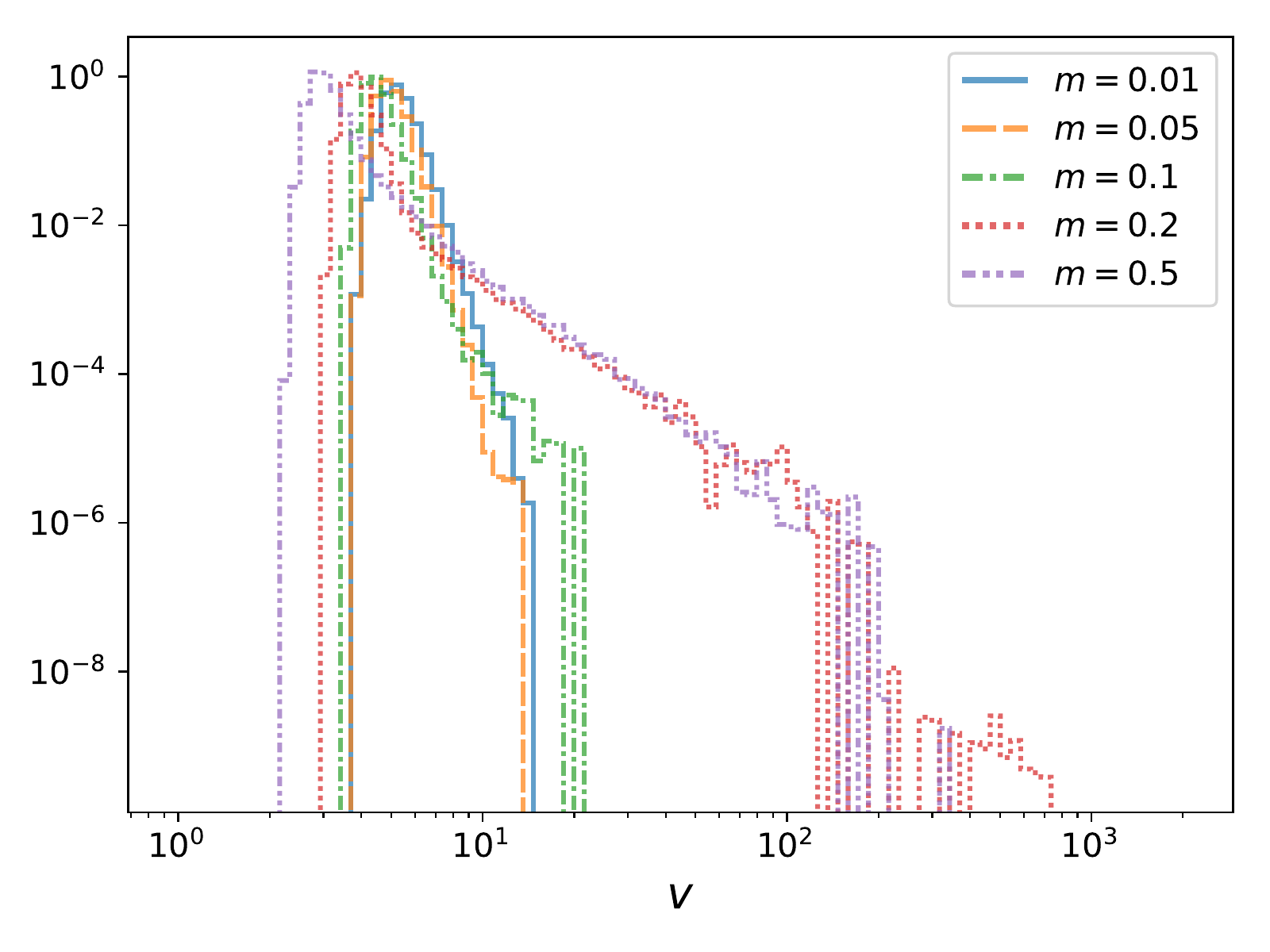}
	\includegraphics[width=7.5cm]{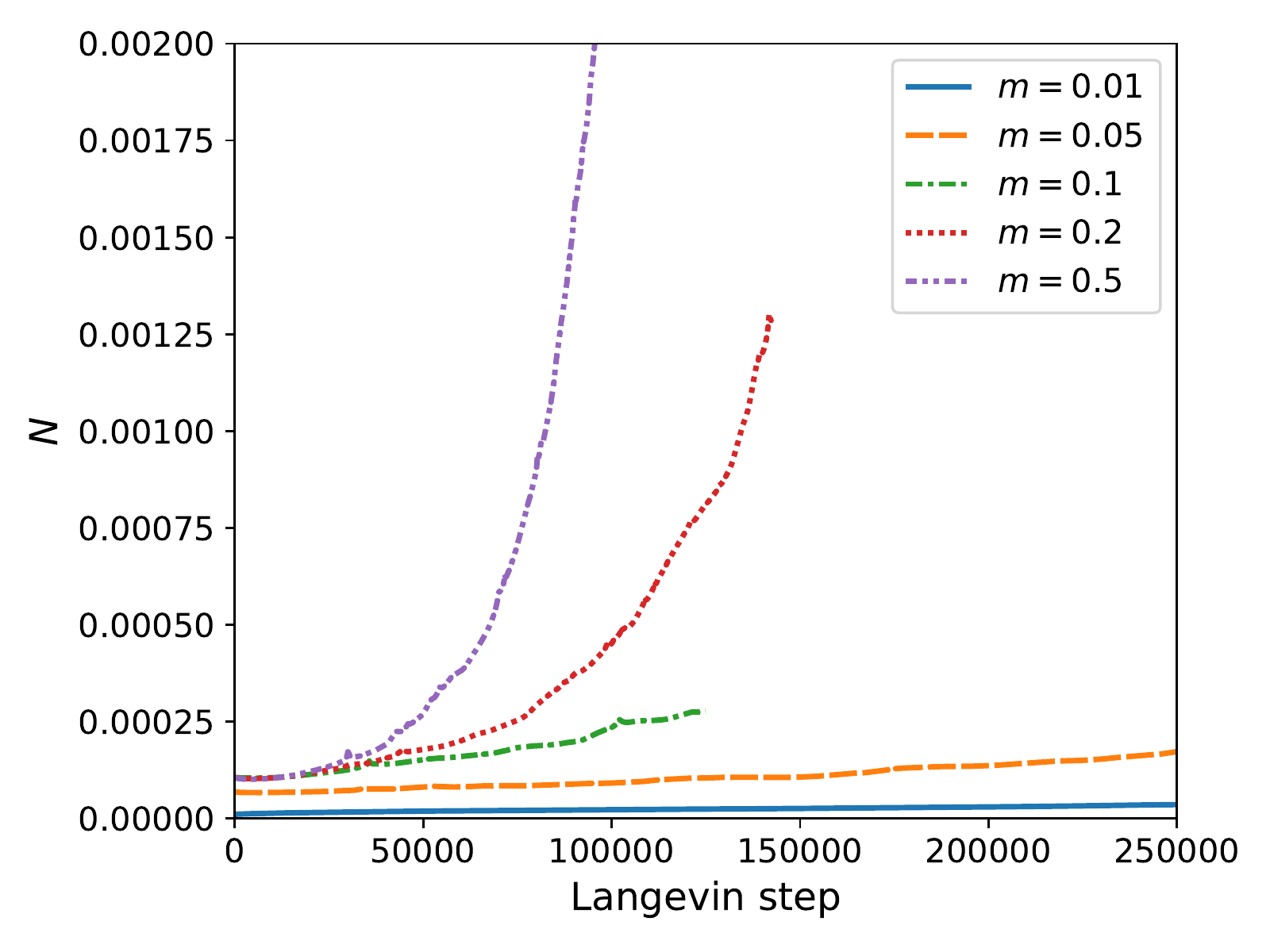}
	\caption{The histogram of the fermionic drift term (Left)
and the history of the unitarity norm (Right)
are plotted for various $0.01 \le m \le 0.5$ with 
$\beta=5.4$ on a $24^3 \times 12$ lattice. 
}
	\label{drift mass dependence}
\end{figure}

In Fig.~\ref{drift mass dependence} we show the probability distribution 
of the drift term and the history of the unitarity norm.
At $m \ge 0.2$,
we find that the distribution of the fermionic drift term
falls off with a power law
and the unitarity norm grows rapidly,
which implies that the CLM is not reliable there.

\begin{figure}
	\centering
	\includegraphics[width=7.5cm]{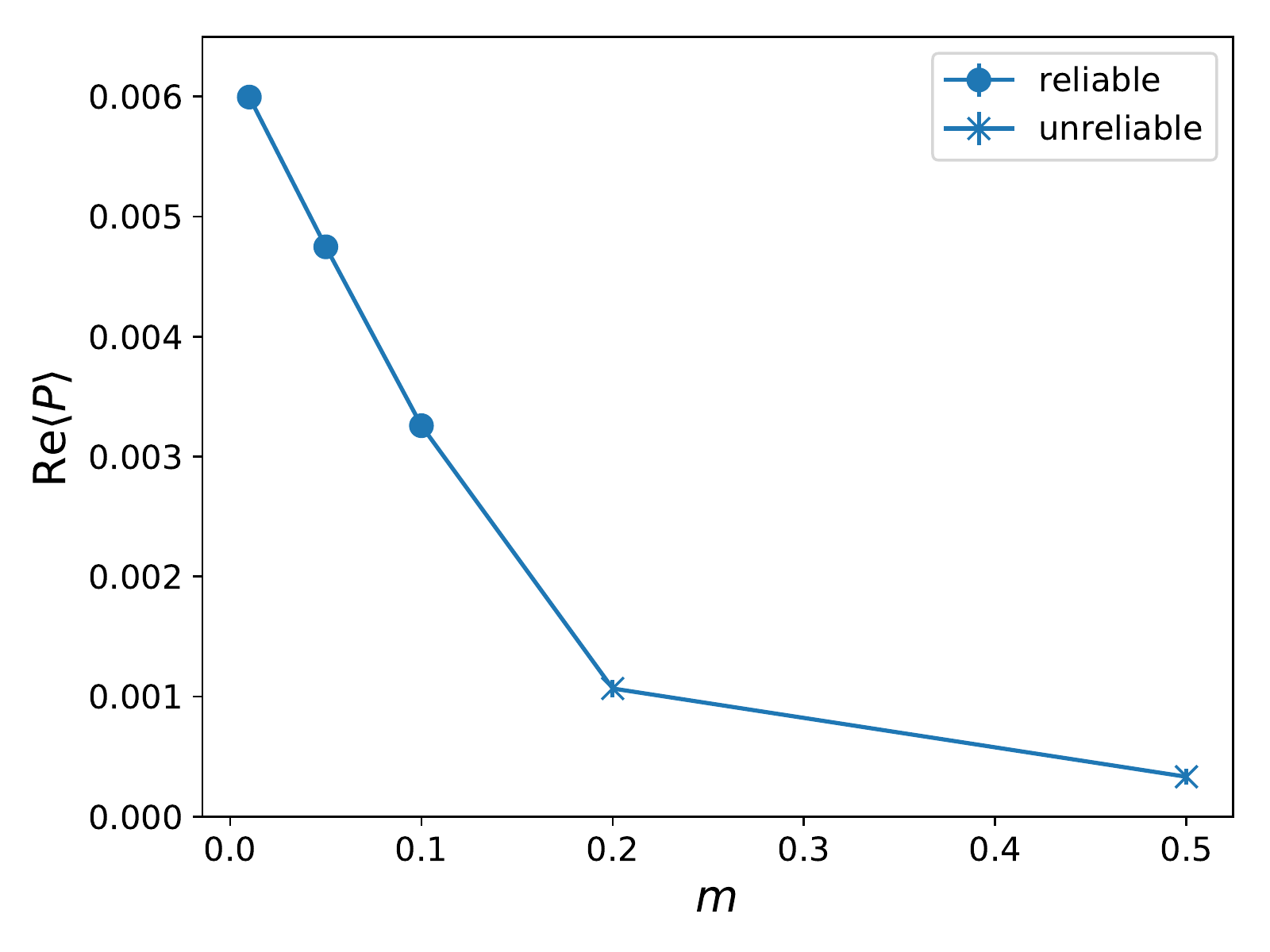}
	\includegraphics[width=7.5cm]{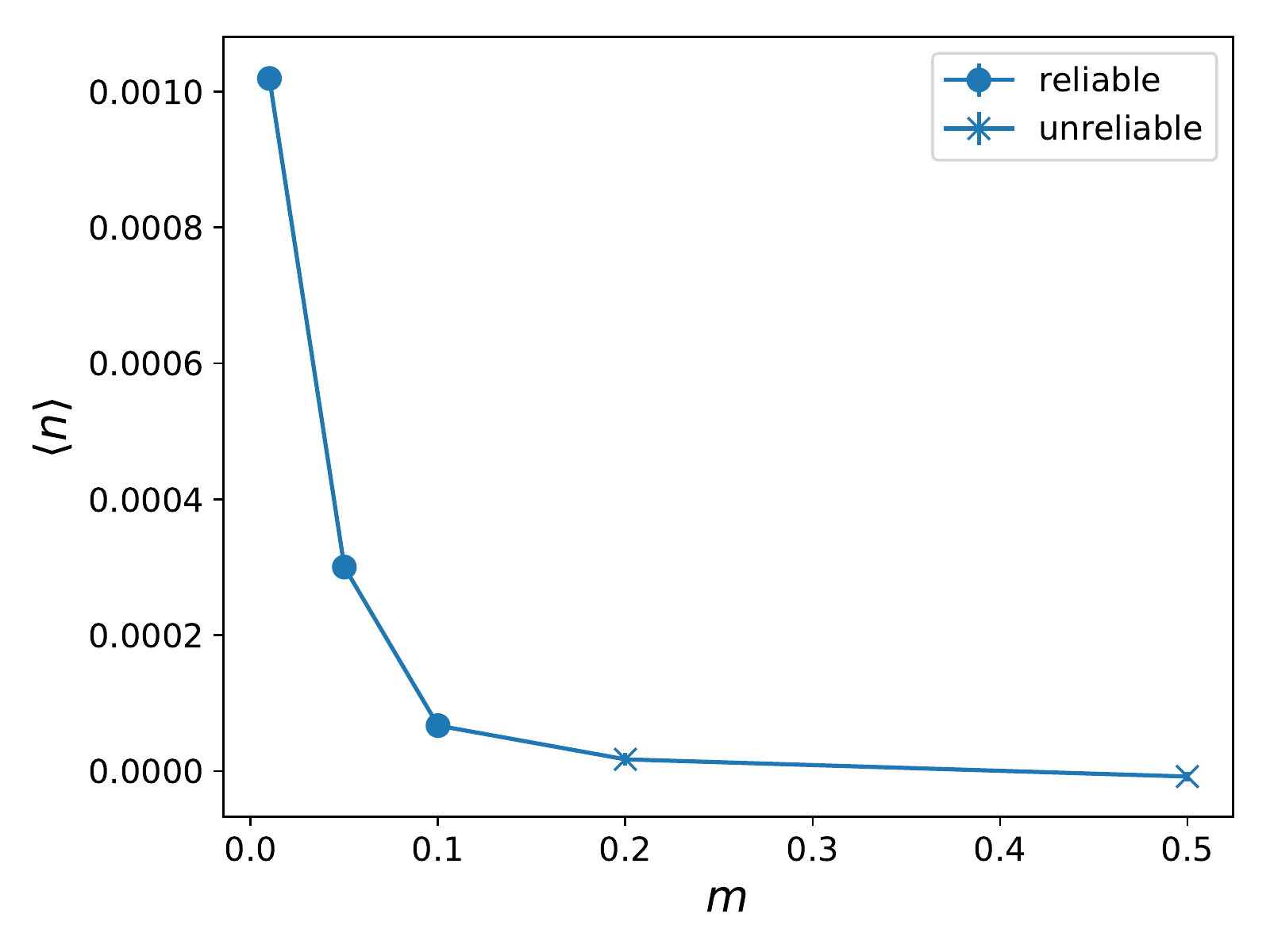}
	\caption{The real part of the Polyakov loop (Left)
and the baryon number density (Right) are plotted against
the quark mass $m$ for $\beta=5.4$ on a $24^3 \times 12$ lattice.
Circles and crosses represent reliable data and unreliable data, respectively.
}
	\label{obs mass dependence}
\end{figure}

In Fig.~\ref{obs mass dependence} 
we show the expectation value of 
the Polyakov loop and the baryon number density
as a function of the quark mass.
Focusing on the reliable data at $m \le 0.1$,
we find that both of them decrease with increasing $m$,
but they remain significantly nonzero suggesting that the system is in 
the deconfined phase.
However, the sharp drop of
the baryon number density at $m \sim 0.1$
suggests that the system enters the confined phase slightly above 
that point.
Thus we find that the first order phase transition is hidden 
in the unreliable region here as well.

\section{Summary and outlook}
\label{summary}

In this paper
we have investigated the validity region of the CLM for lattice QCD 
at finite chemical potential using
the criterion
based on the probability distribution of the drift term.
In particular, we have discussed whether the CLM has an ability 
to probe the deconfinement transition.
We have performed lattice QCD simulations with four-flavor staggered fermions 
on lattices with larger temporal size $N_{\rm t}=12$ than the previous study
with $N_{\rm t}=4,6,8$ so that the critical $\beta$ is shifted to a larger value.
The chemical potential is set to $\mu=0.1$, which corresponds to $\mu/T=1.2$.
Our results obtained at $m=0.01$ on a 
$20^3 \times 12$ lattice
suggest that the singular drift problem
occurs at sufficiently small $\beta$, which seems to hide the phase transition.
As another possibility, we have also increased $m$ at fixed $\beta=5.4$
on a $24^3 \times 12$ lattice, which showed that the singular drift problem
occurs before the phase transition is observed.
When the singular drift problem occurs, the unitarity norm shows a rapid
growth at the same time.
From these results, we speculate that the singular drift problem
occurs in the confined phase quite 
generally\footnote{In ref.~\cite{Nagata:2018mkb}, this problem was avoided 
by using the deformation technique on a $4^3 \times 8$ lattice with $\beta=5.7$,
where the confined phase appears due to finite spatial volume effects
despite the high temperature.} 
unless the quark mass becomes very large.
It would be interesting to see whether the CLM remains applicable
in the deconfined phase even at larger $\mu$ and lower $T$.
Simulations in this direction are underway.

\section*{Acknowledgements}

This research was supported by MEXT as
``Priority Issue on Post-K computer'' 
(Elucidation of the Fundamental Laws and Evolution of the Universe) 
and 
Joint Institute for Computational Fundamental Science (JICFuS).
Computations were carried out
using computational resources of the K computer 
provided by the RIKEN Advanced Institute for Computational Science 
through the HPCI System Research project (Project ID:hp180178).
J.~N.\ was supported in part by Grant-in-Aid 
for Scientific Research (No.\ 16H03988)
from Japan Society for the Promotion of Science. 
S.~S.\ was supported by the MEXT-Supported Program 
for the Strategic Research Foundation at Private Universities 
``Topological Science'' (Grant No.\ S1511006).

\bibliography{lattice2018_tsutsui}
\bibliographystyle{h-physrev5}

%

\end{document}